\documentclass[conference]{IEEEtran}
\IEEEoverridecommandlockouts
\usepackage{cite}
\usepackage{amsmath,amssymb,amsfonts}
\usepackage{mathtools}
\DeclarePairedDelimiter{\ceil}{\lceil}{\rceil}
\usepackage{algorithmic}
\usepackage{graphicx}
\usepackage{subfigure}
\usepackage{textcomp}
\usepackage{xcolor}
\usepackage{comment}

\def\BibTeX{{\rm B\kern-.05em{\sc i\kern-.025em b}\kern-.08em
    T\kern-.1667em\lower.7ex\hbox{E}\kern-.125emX}}
\begin{document}

\title{On the Feasibility of Battery-Less LoRaWAN Communications using Energy Harvesting}


\author{
    \IEEEauthorblockN{Carmen Delgado\IEEEauthorrefmark{1}, Jos\'e Mar\'ia Sanz\IEEEauthorrefmark{2}, Jeroen Famaey\IEEEauthorrefmark{1}}
    \IEEEauthorblockA{\IEEEauthorrefmark{1}IDLab, University of Antwerp -- imec, Antwerp, Belgium
    }
    \IEEEauthorblockA{\IEEEauthorrefmark{2}Kintech Engineering, Zaragoza, Spain
    }
}

\maketitle

\begin{abstract}
From the outset, batteries have been the main power source for the Internet of Things (IoT). However, replacing and disposing of billions of dead batteries per year is costly in terms of maintenance and ecologically irresponsible. Since batteries are one of the greatest threats to a sustainable IoT, battery-less devices are the solution to this problem. These devices run on long-lived capacitors charged using various forms of energy harvesting, which results in intermittent on-off device behaviour. In this work, we model this intermittent battery-less behaviour for LoRaWAN devices. This model allows us to characterize the performance with the aim to determine under which conditions a LoRaWAN device can work without batteries, and how its parameters should be configured. Results show that the reliability directly depends on device configurations (i.e., capacitor size, turn-on voltage threshold), application behaviour (i.e., transmission interval, packet size) and environmental conditions (i.e., energy harvesting rate).

\end{abstract}


\section{Introduction}\label{sec:intro}

In the Internet of Things (IoT) vision, billions of interconnected devices cooperate to sense, actuate, locate and communicate with each other with the aim of supporting and improving daily life. Recent advancements in ultra-low power communications technologies are driving the transformation of everyday objects into an information source connected to the Internet.  Usually, these devices are equipped with a battery, a radio chip, a microcontroller unit (MCU) and one or more sensors and/or actuators.

Low-Power Wide-Area Networks (LPWANs) are a new set of radio technologies that are designed to support the needs of IoT deployments by combining low energy consumption with long range communications \cite{Raza2017}. LoRaWAN \cite{LoraWan} is an LPWAN technology that uses sub-GHz unlicensed spectrum and enables long-range transmissions (more than 10km in rural areas) at low power consumption~\cite{Sanchez2018}.

However, from the beginning, battery limitations have been one of the main problems to realise the vision of a global IoT. Sometimes, devices are located in hard-to-reach areas, and battery replacement or maintenance is not only costly, but also dangerous. 
In other situations, small devices are needed, but batteries are too bulky. In general terms, it is known that batteries are expensive, bulky and hazardous; they are sensitive to temperature, short-lived, and therefore incompatible with a sustainable IoT~\cite{Hester2017}.

Since the number of IoT devices is quite high and will continue to increase in the coming years, it seems clear that the use of batteries should be reconsidered. 
To alleviate the IoT's battery problem, battery-less IoT devices are a promising solution. 
Battery-less IoT devices are smaller, live longer, are more environmentally friendly, and are cheaper to maintain.
This makes them especially suitable for applications in hard-to-reach locations (e.g., intra-body health monitoring, remote-area sensing) and large-scale deployments (e.g., dense building automation networks, smart cities).

\begin{figure}[t]
\centerline{\includegraphics[width=1\columnwidth]{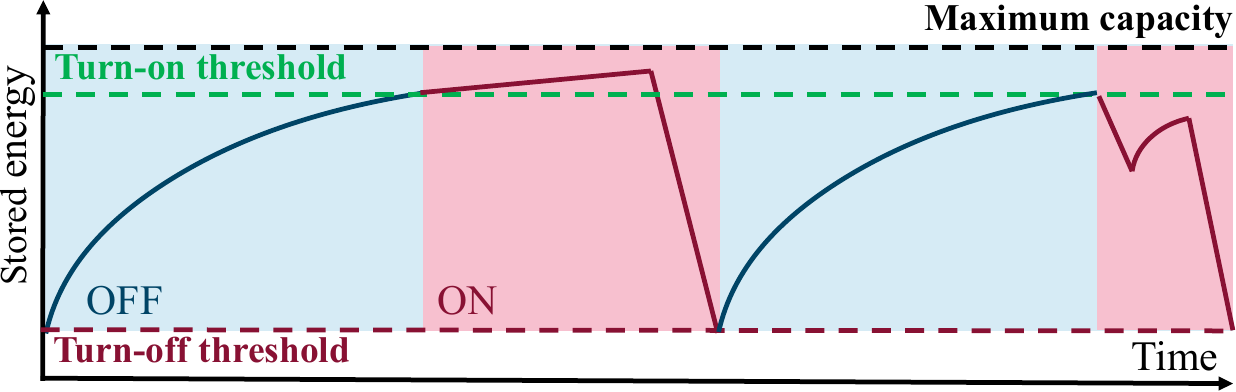}}
\caption{Turn-off and turn-on threshold of battery-less intermittent behaviour}
\label{fig:behaviour}
\end{figure}

In contrast to their battery-powered counterparts, battery-less
devices need to harvest energy from their environment and store it in tiny capacitors. Such capacitors easily last more than ten years \cite{Spanik2014} thanks to the fact that they can handle a larger number of charge cycles. Moreover, they are cheaper to produce and easy to recycle, thus better for the environment. However, this new paradigm faces some difficulties: energy harvesting is inconsistent, energy storage is scarce, power failures are inevitable, and execution is intermittent \cite{Hester2017}. This intermittency, illustrated in Figure~\ref{fig:behaviour}, causes the device to turn on and off frequently, as it swiftly depletes the energy stored in the capacitor. This results in a power failure until the device harvests enough energy to turn on again.

In this paper, we propose a system model for a battery-less LoRaWAN Class A device. This allows us to characterize device performance in terms of reliability for various device configurations (e.g., capacitor size, transmission interval, packet size, turn-on voltage threshold) and environmental conditions (e.g., energy harvesting rate). Based on this analysis, we determine the feasibility of using LoRa without batteries.

\section{Related Work}\label{sec:relatedwork}
One of the main problems of IoT devices is their energy. When the energy of a device is depleted, it will no longer fulfill its role unless either the source of energy is replaced or a harvesting mechanism is used. Since we posit the vision of battery-less devices, energy harvesting is the only way to overcome the energy gap. It is a mechanism that allows extracting energy from external sources, such as solar irradiance, wind, thermoelectric, piezoelectric, or vibration~\cite{Shaikh2016}. 

Sherazi \textit{et al.} present a mathematical model to calculate the battery life using energy harvesting in order to analyze the impact of energy harvesting sources~\cite{Sherazi2018}. Although they also use LoRaWAN in their calculations, they do assume the use of batteries.
In contrast, we consider that the harvested energy is saved in a capacitor, which has a significantly lower energy storage capacity. The smaller its capacity, the faster it will charge and discharge for a given harvester, which can lead to power failures up to several times per second~\cite{Ransford2011}.  

Most prior works on battery-less devices use passive RF-powered communications, such as backscatter \cite{Talla2017} \cite{Correia2016}. Instead, we focus on active communications. 
Only a few works have considered the combination of battery-less devices, energy harvesting and active radios. For example, Fraternali \textit{et al.} designed a battery-less Bluetooth Low Power (BLE) sensor node which leverages ambient light and a power management algorithm to maximize the quality of service \cite{Fraternali2018}. 
Jornet and Akyildiz present an energy model for nanodevices using energy harvesting mechanisms~\cite{Jornet2012}. In their model, they also consider energy harvesting with a capacitor, and then they analyze the energy consumption of the communications in the Teraherz band.
However, the communications and networking solutions considered at nanoscale are vastly different than those in traditional IoT systems, hence leading to a significantly different system model.

\section{Model for a battery-less LoRaWAN device}\label{sec:model}

\subsection{Circuit Characterization}\label{sec:circuitmodel}

\begin{figure}[t]
\centerline{\includegraphics[width=0.7\columnwidth]{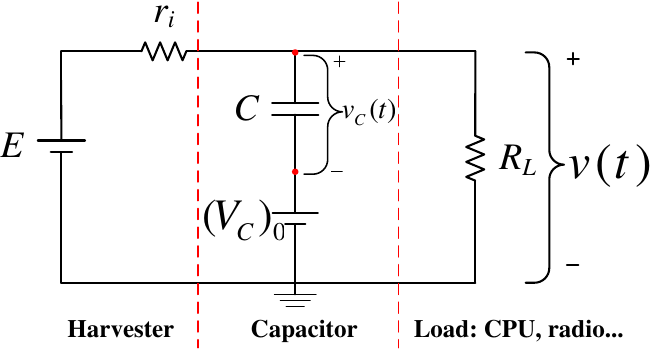}}
\caption{Electrical circuit model of a battery-less IoT device }
\label{fig:Circuit}
\end{figure}

Battery-less IoT devices are equipped with a harvester mechanism, a capacitor, an MCU, a radio unit and the needed sensors. In order to model the behaviour of these devices, we have considered the simplified electrical circuit shown in Figure~\ref{fig:Circuit}, where the circuit has been divided into three main parts: the harvester (source of the energy), the capacitor (storage of the energy) and the load (consumer of the energy: MCU, radio, sensors). More details about these 3 parts are explained below.

\paragraph{Energy Harvesting System Model} 

Although in reality the harvester model should depend on the type of energy source (e.g., radiation, vibration), 
we consider a generic and simplified approach, where only the generated power is taken into account.
As can be seen in the left side of Figure~\ref{fig:Circuit}, the harvester is modelled as a real voltage source composed of an ideal voltage source and a series resistance (denoted by $E$ and $r_i$, respectively).
The value of $E$ (in Volts) is chosen according to the operating voltage of the circuit elements, which in this case will be determined by the load.
The series resistance $r_i$ limits the power of the harvester and its value is calculated using the following equation: 
\begin{equation}
 r_i = \frac{E^2}{P_{harvester}}
\end{equation} \label{eq:ri} 
where $P_{harvester}$ is the power of the harvester source, which can vary greatly depending on the type of energy harvesting considered (e.g., up to $1 mW/cm^2$ for indoor natural light, and up to $100 mW/cm^2$ for outdoor sun) \cite{Shirvanimoghaddam2018}. 

\paragraph{Capacitor Model} 
The capacitor is the part of the circuit where the energy is stored. As shown in Figure~\ref{fig:behaviour}, the behaviour of the system is a succession of intervals, where the capacitor is being charged or discharged. Each interval is characterized by a specific state of the load components (e.g., MCU is active and radio is transmitting). 
We characterize the voltage of the capacitor throughout each interval using $(V_C)_0$ and $v_C(t)$. $(V_C)_0$ represents the initial voltage of the capacitor at the beginning of the interval (i.e., time $t_0$), and $v_C(t)$ is the temporal evolution of the said voltage at time $t$ (relative to $t_0$). Both $(V_C)_0$ and $v_C(t)$ are included in the circuit as an ideal voltage source and the voltage over time of an ideal capacitor respectively, as shown in Figure~\ref{fig:Circuit}.

\paragraph{Load Model} The load of the model corresponds to the set of components that consume the stored energy in the capacitor, such as the MCU, radio or sensors. Each of these components is characterized by a specific power consumption in each of its states (e.g., active, sleeping, off).
Therefore, they can be modelled as a load resistance denoted by $R_L$ (in $\Omega$), which can be calculated as follows:
\begin{equation}
 R_L = \frac{E}{I_{load}}
 \label{eq:RL} 
\end{equation} 
where $I_{load}$ is the sum of the supply currents of all components in their current state. $R_L$ thus varies across intervals depending on the state of each component during that interval. 

To determine if the device has enough energy at a specific time to perform its tasks (e.g., transmit data), it is needed to calculate the voltage across the load of the model $v(t)$:
\begin{equation}
 v(t) = E\frac{R_{eq}}{r_i}(1-e^{(\frac{-t}{R_{eq}C})})+(V_C)_0e^{(\frac{-t}{R_{eq}C})}
 \label{eq:V} 
\end{equation} 
where $C$ is the capacitance in Farads, $t$ is the time spent in the current interval,
and $R_{eq}$ is the equivalent resistance of the circuit (in $\Omega$), computed as:
\begin{equation}
 R_{eq}=\frac{R_L r_i}{R_L + r_i}
 \label{eq:Req} 
\end{equation} 

The value of $v(t)$ provides the voltage available in the load, which will be used to determine if a specific action (e.g., transmit, listen) can be performed during an interval, according to the needed time $t$ it will take, the energy harvesting rate $P_{harvester}$, the specific load $I_{load}$, and the capacitor voltage $(V_C)_0$ at the start of the interval. Note that $v(t)$ can be increasing or decreasing depending of the specific parameters.

\subsection{LoRAWAN Class A Device Model}\label{sec:systemmodel}

We consider a LoRa device, using the LoRaWAN medium access control (MAC) protocol. The LoRaWAN standard defines three classes of end-devices. In this work, we focus on class A, which provides the lowest energy consumption~\cite{LoraWan}.

\begin{figure}[t]
\centerline{\includegraphics[width=1\columnwidth]{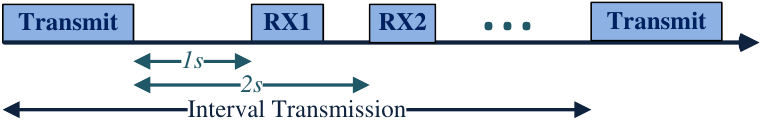}}
\caption{LoRaWAN Class A end device window timings}
\label{fig:Lora}
\end{figure}

Class A devices spend most of the time in deep sleep, and only wake up when they need to transmit data to the network server. Since they use an ALOHA-based MAC protocol, Class A LoRAWAN devices do not perform listen before talk. It is also characteristic that devices of this class are only reachable for downlink transmissions after they transmit. As can be seen in Figure~\ref{fig:Lora}, LoRaWAN class A devices have two reception windows.
After the transmission, the device waits 1 second in idle state and then switches to listening mode (RX1). If the device receives a downlink transmission in this first window, it does not need to stay awake for the second reception window. However, if nothing is received, it must switch again to idle mode. In that case, 2~seconds after the end of the transmission, the device will listen again (RX2).

One of the important parameters to be configured 
is the spreading factor (SF), which represents the ratio between the chip rate and the baseband information rate and can range from 7 to 12. As the SF increases, so does the coverage range and decoding robustness. This comes at the cost of a decrease in data rate, and thus increase in airtime. In fact, the SF determines the TX time as well as the RX1 time. 
The time of RX2 is fixed, as it always uses the slowest spreading factor SF12.
The formulas for calculating the different times can be derived from the LoRaWAN standard~\cite{LoraWan}. First, it is convenient to define the symbol duration $T_{sym}$ (considering SF bits per symbol):
\begin{equation}
 T_{sym}=\frac{2^{SF}}{BW}
 \label{eq:Tsym} 
\end{equation} 
where $BW$ is the bandwidth, which is typically 125kHz.

\begin{figure}[t]
\centerline{\includegraphics[width=0.9\columnwidth]{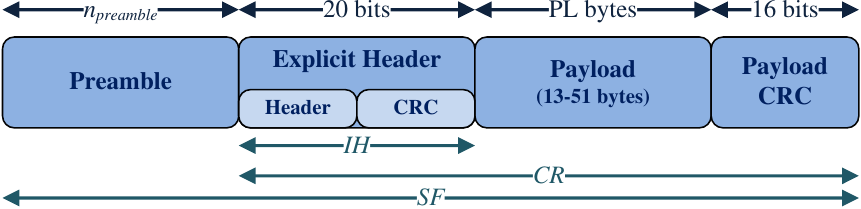}}
\caption{LoRaWAN packet formatting}
\label{fig:lora_packet}
\end{figure}

Figure~\ref{fig:lora_packet} shows the LoRa frame format of the physical layer. The preamble ($n_{preamble}$) is the number of programmed preamble symbols, which is 8 in LoRaWAN 1.0. Then, the low-level header can be explicitly enabled ($IH=0$) or disabled ($IH=1$). It is used to indicate the coding rate and payload length, and can be left out if both sides of the communication have these parameters fixed. The time needed to transmit or receive the preamble sequence is given by:
\begin{equation}
 T_{preamble}=(n_{preamble} + 4.25) \cdot T_{sym}
 \label{eq:Tpreamb} 
\end{equation} 

The number of symbols that make up the packet payload and header is given by Equation~\ref{eq:PLSym}, and the payload duration is given by Equation~\ref{eq:Tpayload}.
\begin{multline}
S_{payload} = 8 + \\
 max (\ceil[\Bigg]{ \frac{8PL - 4SF + 28 +16 - 20 IH}{4(SF - 2DE)}} (CR + 4), 0)
 \label{eq:PLSym} 
\end{multline} 
\begin{equation}
 T_{payload} = S_{payload} \cdot T_{sym}
 \label{eq:Tpayload} 
\end{equation}
where $PL$ is the number of payload bytes (which can vary from 13 to 51), $CR$ is the coding rate where higher values mean more overhead, and the low data rate optimization can be enabled with $DE =1$ and disabled with $DE=0$ (which intends to correct the clock drift at SF11 and SF12). 
Finally, the time on air (or packet duration) can be calculated as follows:

\begin{equation}
 T_{packet} = T_{preamble} + T_{payload}
 \label{eq:Tpacket} 
\end{equation} 

\subsection{Complete System Model}\label{sec:CompleteSystemModel}
The main problem of battery-less devices is how to deal with the intermittent behaviour, and the energy and time needed to be awake to perform the different actions (e.g., listen, transmit, receive). For simplicity of the analysis, we assume a constant energy harvesting rate during a single experiment.
As shown in Figure~\ref{fig:behaviour}, an intermittent device has two main states: ON and OFF. When the device is in the ON state, it can be in one of the following sub-states: Sleep, Idle, TX, Listen, or RX. In these cases, it will stay turned on until it actively turns itself off, or when the capacitor voltage drops below the turn-off threshold $V_{min}$. This is the lowest voltage at which its radio and MCU can safely operate. When the device is in the OFF state, it will stay turned off until its capacitor reaches the turn-on voltage threshold $V_{SL}$, which is a configurable parameter. 
Depending on the $P_{harvester}$ and the power consumption of the MCU (which determines $R_L$), the capacitor can be charging or discharging during sleep mode (cf., Equation~\ref{eq:V}).

\begin{figure}[t]
\centerline{\includegraphics[width=0.9\columnwidth]{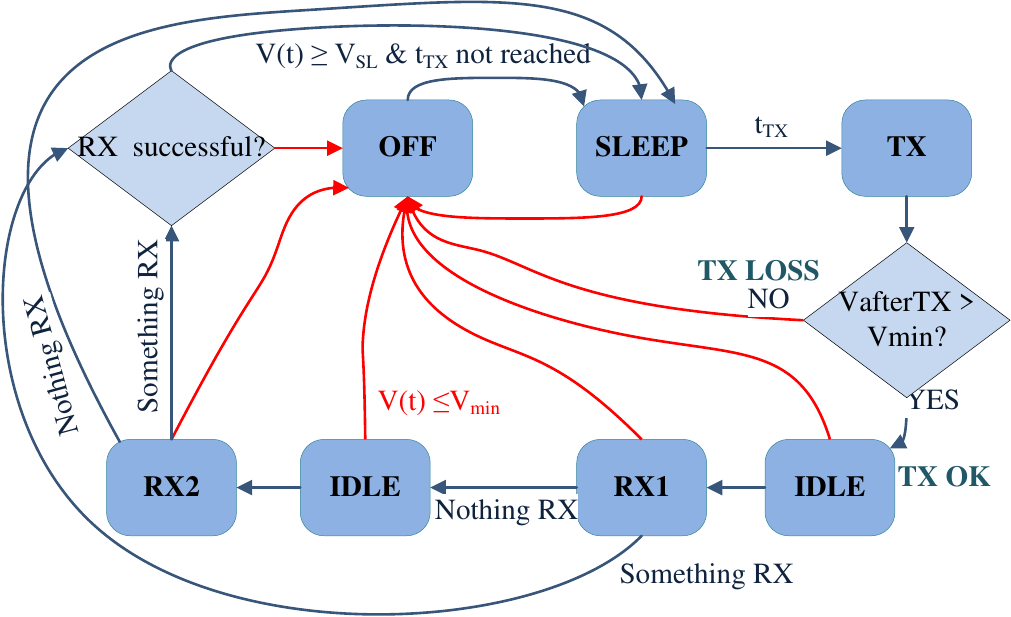}}
\caption{State diagram of the complete system model}
\label{fig:blocks}
\end{figure}

The complete state diagram of the system is shown in Figure~\ref{fig:blocks}.  Whenever the device has something to transmit, it will be checked in which state the system is. If the system is in off mode, it will not be able to transmit and the packet will be lost. However, if the system is in the sleep state, it will try to send it. If the current voltage in the load is enough to transmit it, it will do it and will switch to idle state for 1 second (as shown in Figure~\ref{fig:Lora}). In order to know the needed voltage to transmit, Equation~\ref{eq:V} is used, where the time $t$ corresponds to the time on air (or time needed to send the packet) that can be calculated with Equation~\ref{eq:Tpacket}. After being 1 second in the idle state, the system will switch to listen, in the first reception window (RX1), where the device checks whether a preamble has been received. If a preamble has been detected, the device continues receiving the downstream transmission, and after that it will switch to sleep mode. The maximum time spent in the first listening window is determined by Equation~\ref{eq:Tpreamb}, where the $SF$ is the corresponding spreading factor used in the transmission. If a preamble has not been detected during this time, it closes the receive window and transitions to the second idle state. 2 seconds after the end of the transmission, the system will switch to the second reception window (RX2), where the device will check again if a preamble has been received. In this case, the maximum time spent in this window is determined by Equation~\ref{eq:Tpreamb} but using a $SF$ value of 12.
The system will switch to the off state immediately whenever the voltage of the capacitor reaches the $V_{min}$ value, and any ongoing transmission or reception will fail.

\section{Analysis of the model}\label{sec:analysis}

In this section, we evaluate the performance of a battery-less LoRaWAN Class A device in terms of reliability for various device configurations and environmental conditions. This evaluation provides an overview of the needed parameters and elements that will allow a battery-less LoRaWAN device to work for uplink transmissions. In this analysis, we do not consider downlink transmissions (although we do consider empty RX1 and RX2 slots), the energy consumed by sensors (which in many cases is negligible compared to energy consumed by the radio), or collisions between devices. These aspects are left for future work.

We first provide an analysis of the needed capacitor and the minimum needed voltage. Then, we evaluate the impact of the turn-on voltage threshold on the reliability in terms of packet delivery ratio (PDR) for different data transmission rates.

\subsection{Simulation Setup}

The simulation parameters are based on the Semtech SX1272/73 LoRa radio. . 
As such, $V_{min}$ and $E$ have been defined as 1.8V (minimum operating voltage of the SX1272/73) and 3.3V (typical operating voltage), respectively.
For all the cases, we assume periodic uplink transmissions at a constant interval. Table~\ref{tab:generalparameters} summarizes the general parameters used in the simulations.

\begin{table}[t]
\begin{footnotesize}
\caption{General simulation parameters}
\begin{center}
\begin{tabular}{l l r}
\hline
\textbf{Parameter}& \textbf{Symbol}& \textbf{Value} \\
\hline
min Voltage & $V_{min}$ & 1.8V \\
max Voltage of the system & $E$ & 3.3V \\
Coding Rate & $CR$ & 4/5 \\
Bandwidth & $BW$ & 125kHz \\
Preamble symbols & $n_{preamble}$ & 8\\
Data rate optimization enabled & $DE$ & 0\\
Head disabled & $IH$ & 1\\
\hline
\end{tabular}
\label{tab:generalparameters}
\end{center}
\end{footnotesize}
\end{table}

As explained before, the values used for the load depend on the specific state of the load components. We consider that the load is formed by the radio and the MCU (i.e., an STM32L162xE chip. 
In this case, current consumption in  Low-power run mode (active) and in Low-power sleep mode (sleep) are considered. Table~\ref{tab:simStates} defines the considerations of the MCU and the radio for the different states of the system and the corresponding values for the load, $R_L$, which are determined using Equation~\ref{eq:RL}.

\begin{table}[t]
\begin{footnotesize}
\caption{States of the system}
\begin{center}
\begin{tabular}{l c c r}
\hline
\textbf{System State}& \textbf{MCU State}& \textbf{Radio State} & \textbf{$R_L$} \\
\hline
Off & Off &  Off  &  600 $k\Omega$ \\
Sleep & Sleep & Sleep  &  589.286 $k\Omega$\\
Idle & Sleep & Idle  &  471.428 $k\Omega$\\
Tx (+13 dBm) & Active &  TX & 117.811 $\Omega$ \\
Listen & Active & Listen &  313.957 $\Omega$ \\
Rx & Active & RX  &  294.354 $\Omega$\\

\hline
\end{tabular}
\label{tab:simStates}
\end{center}
\end{footnotesize}
\end{table}

\subsection{Analysis of the Capacitance and Voltage Requirements}

The Class A LoRaWAN standard defines the sequence to follow when performing a transmission as mentioned in Section~\ref{sec:systemmodel}. In order to characterize the system, it would be useful to know what kind of capacitor is needed to support it.

\begin{figure}[t]
\centerline{\includegraphics[width=0.8\columnwidth]{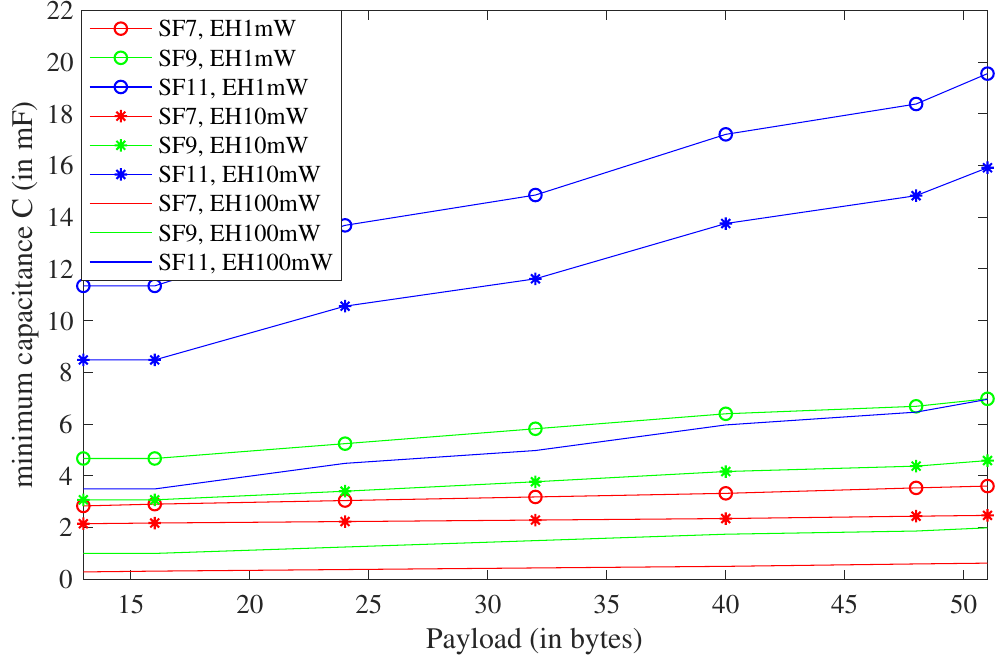}}
\caption{Minimum capacitance needed when varying the payload}
\label{fig:minC}
\end{figure}

Figure~\ref{fig:minC} shows the minimum capacitance ($C$) needed to complete one transmission cycle for different values of payload, SF, and energy harvesting. Such a cycle includes transmitting a packet and staying awake for RX1 and RX2 (assuming nothing is received). 
For high data rates (i.e., small SF values), the payload has little influence on required capacitance. However, when using SF11, the impact of the packet size is considerable (e.g., varying from $11000\mu{}F$ to $19500\mu{}F$ when the energy harvesting rate is set to $1mW$). As expected, lower energy harvesting rates are not compatible with larger spreading factors, as the required capacitance becomes too high, which is not compatible with the low cost and small form factor requirements of an IoT device.

Let us consider a capacitor with $C = 4700\mu{}F$, which is in line with the capacitance of low-cost off-the-shelf Aluminum-Polymer capacitors. The figure shows that such a capacitor could support SF7 at any packet size, and SF9 for small packet sizes (i.e., 16 bytes or less), even at an energy harvesting rate of 1mW.

In order to get a deep insight in the specific values, Table~\ref{tab:minV} shows the minimum voltage needed to perform a complete a transmission cycle for such a capacitor. The energy harvesting rate $P_{harvester}$ is represented as EH. According to these results, we can see that SF is the parameter with the most impact. This is due to the fact that it greatly affects the time needed to transmit, so that, more (or less) energy is used. On the other hand, the influence of the payload size is small for SF7, but increases with increasing SF and decreasing EH.
The reason higher energy harvesting rates reduce the effect of the payload size is because the loss of energy when performing device actions is mitigated while harvesting. Additionally, if the data rate is high (and SF low), larger packet sizes have a negligible effect on the time-on-air. While for some configurations it is possible to complete a cycle, for others it is impossible (marked as -). These cases correspond to those that show a higher minimum capacitance than $4700\mu{}F$ in  Figure~\ref{fig:minC}. For this reason, it will be needed to choose the proper capacitor size according to the specific environmental conditions (energy harvesting rate) and network conditions (SF and payload size).

\begin{table}[t]
\begin{scriptsize}
\begin{center}
\caption{Voltage needed to complete a transmission cycle}
\begin{tabular}{ccccccc}
\hline
  &   & \multicolumn{5}{c}{\textbf{Payload (bytes)}}     \\
 \cline{3-7}
   \textbf{EH}       &    \textbf{SF}    & \textbf{16}     & \textbf{24}     & \textbf{32}     & \textbf{40}     & \textbf{48}        \\
           \hline
        								& \textbf{7}   & 2.5628 & 2.6105 & 2.6591 & 2.7086 & 2.7846  \\
   \textbf{1mW }        		& \textbf{9}   & 3.228  & - & - & - & -   \\
           							& \textbf{11}    & -      & -      & -      & -      & -           \\
             \hline
        								& \textbf{7}  & 1.9604 & 1.9943 & 2.0289 & 2.0643 & 2.1186  \\
 \textbf{10mW}           	& \textbf{9} & 2.4184 & 2.5975 & 2.7919 & 3.0028 & 3.115  \\
            							& \textbf{11}     & -      & -      & -      & -      & -           \\
             \hline
       								& \textbf{7}  & 1.8183 & 1.8224 & 1.8266 & 1.8309 & 1.8378 \\
\textbf{100mW}            	& \textbf{9} & 1.8732 & 1.8995 & 1.9302 & 1.9661 & 1.9862  \\
            							& \textbf{11}& 2.4617 & 3.0976 & -      & -      & -          \\
             \hline
\end{tabular}
\label{tab:minV}
\end{center}
\end{scriptsize}
\end{table}

\begin{figure*}[t]
	\centering
	{
		\subfigure[SF7, Payload 16B, EH 1mW]{\includegraphics[width=0.65\columnwidth]{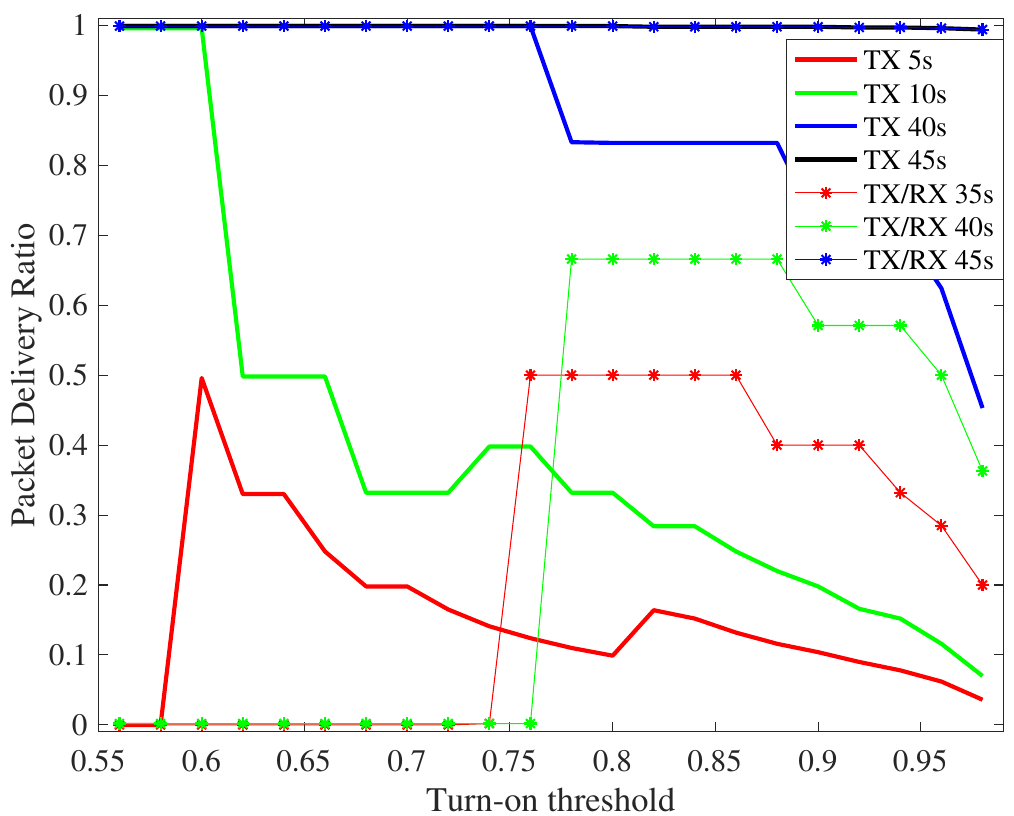}
			\label{fig:SF7_PS16_EH1}}
		\subfigure[SF7, Payload 48B, EH 1mW]{\includegraphics[width=0.65\columnwidth]{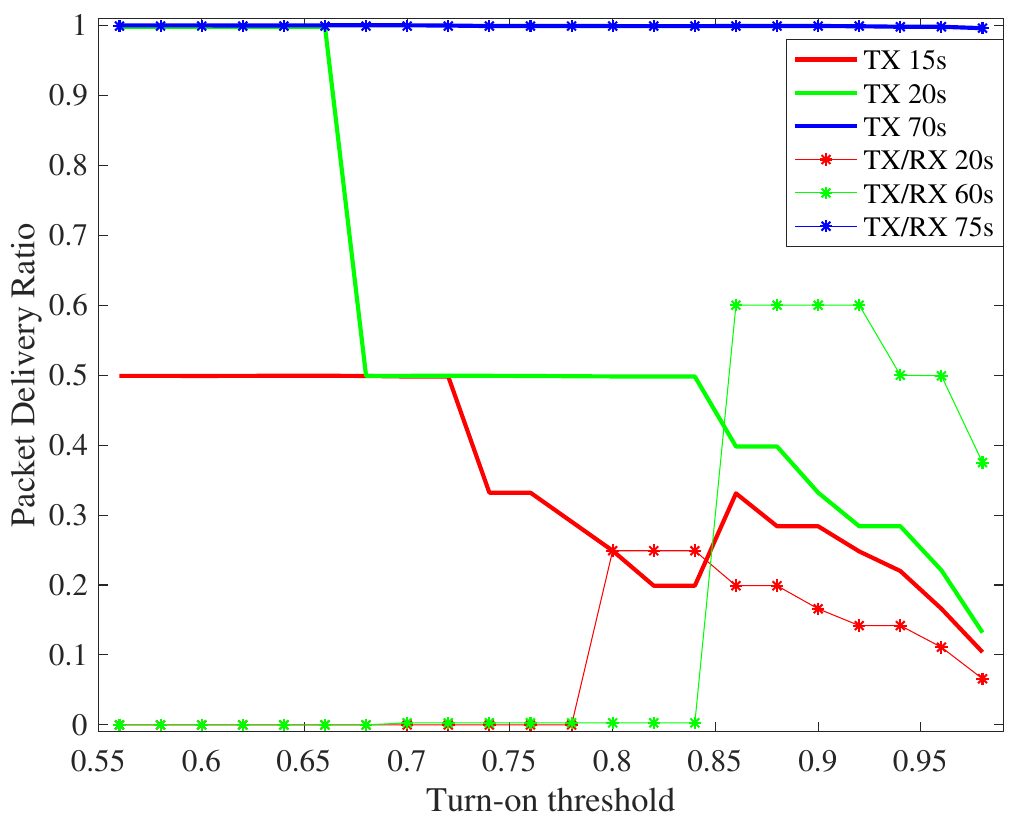}
			\label{fig:TX_SF7_PS48_EH1}}
		\subfigure[SF9, Payload 16B, EH 1mW]{\includegraphics[width=0.65\columnwidth]{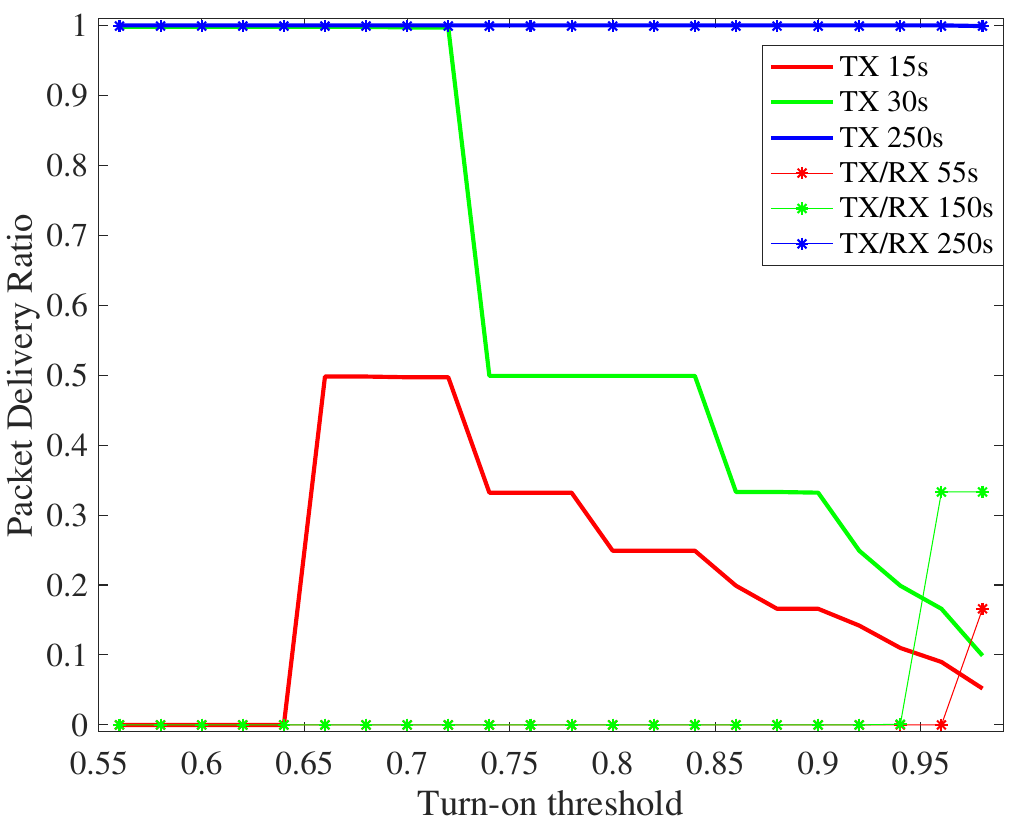}
			\label{fig:TX_SF9_PS16_EH1}}
		
		\caption{Successful TX and TX/RX transmission when varying the wake-up threshold}
		\label{fig:wakeupthreshold}
	}
	\end{figure*}

\subsection{Analysis of the Turn-On Threshold}

Now that we know how to properly select the capacitance, we evaluate how the performance is affected by varying the turn-on voltage threshold parameter. In the following, we consider this threshold as a percentage of the maximum voltage of the system ($E$). We vary the turn on threshold from 55\% (i.e., slightly higher than the 1.8V turn-off threshold) up to 98\%.

Figure~\ref{fig:wakeupthreshold} shows the packet delivery ratio, averaged over 1000 transmissions, as a function of the turn on threshold and for different transmission intervals. These results consider a harvesting rate of 1mW (most realistic for indoor applications), and a capacitance of $4700\mu{}F$. The graphs show both the number of successful uplink transmissions (referred to as TX in the legends), as well as the number of successfully completed cycles (referred to as TX/RX). The latter requires the device to stay awake for the reception windows RX1 and RX2, as it should according to the LoRaWAN standard. The figure shows that if only considering TX, for SF7 the device can successfully transmit every 10 or 20 seconds for a 16 or 48 byte payload respectively. If also required to listen to the two RX opportunities, this increases to 45 and 75 seconds respectively. For SF9, efficiency is greatly reduced, and a 100\% success rate can only be achieved with a small payload size of 16 bytes and a transmission interval of 30 seconds (TX only) or 250 seconds (RX/TX).

When increasing the energy harvesting power ($P_{harvester}$), better performance is achieved. For example, for $P_{harvester} =$ 100 mW, no differences are found when varying the payload size because the time needed to wake up is really small (i.e., 0.40875s when using a wake up threshold of 55\% and 1.53481s when using 95\%). In these scenarios, the device is ON most of the time, and can harvest enough energy to transmit successfully whenever it's turned on, independent of the SF or payload size (as long as it is supported by the capacitor).

\begin{figure}[t]
\centerline{\includegraphics[width=0.8\columnwidth]{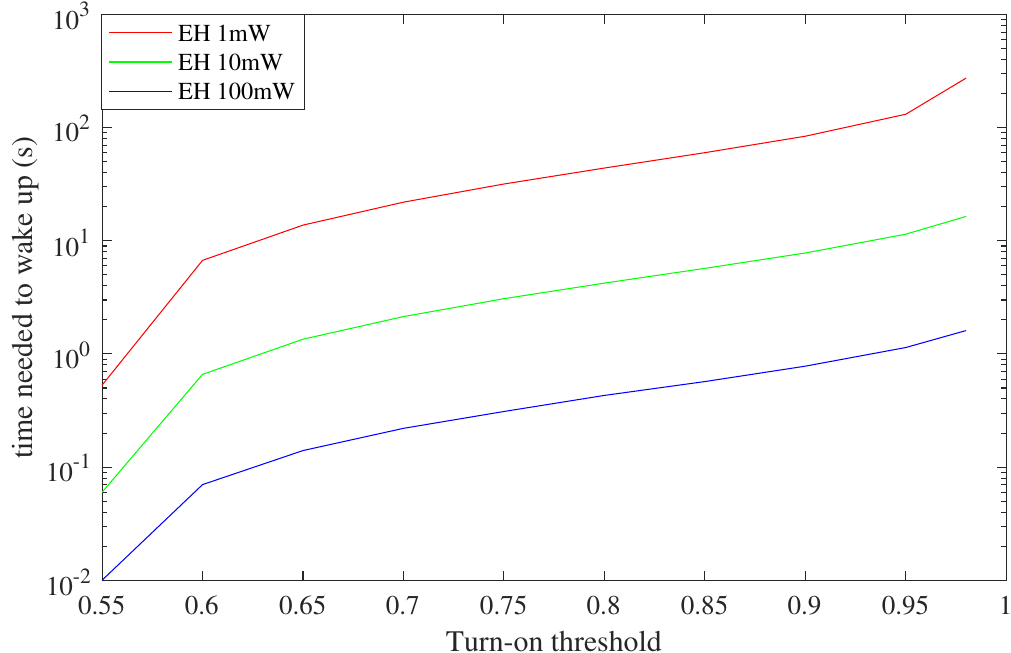}}
\caption{Needed time for turning on for different wake up thresholds}
\label{fig:timetowakeuplog}
\end{figure}

This is illustrated in Figure~\ref{fig:timetowakeuplog}, where the time needed to reach the wake up threshold, starting from $V_{min}$, is shown. This figure shows that for a threshold of 55\%, the device can always wake up within 1 second of turning off. However, when the harvesting rate is low (i.e., 1mW), the time needed to reach the turn on threshold quickly goes above 10 seconds for 65\%, and up to 120 seconds for 98\%. Based on this, it is clear that a low turn-on threshold allows the device to transmit more often, but it may try to transmit when it does not have enough stored energy to do so. A high turn-on threshold, causes much longer off intervals, causing the device to potentially miss transmission opportunities, but ensuring it has enough energy to perform its actions once it wakes up. This is also illustrated in Figure~\ref{fig:wakeupthreshold}. In some cases (e.g., SF7, 16 byte payload, 1mW harvesting, TX only every 10 seconds) the PDR is maximized for low turn on thresholds close to 55\%. In other cases (e.g., SF7, 16 byte payload, 1mW harvesting, TX/RX every 40 seconds) the optimal turn on threshold is higher, and both very low and very high thresholds lead to reduced PDR. In general, operations that require little energy (relative to the energy harvesting rate and capacitance) benefit from a lower turn on threshold. As the relative energy required to complete the operation increases, so does the optimal turn on threshold. This clearly shows the need for intelligent turn on mechanisms that take into account the environment, device and network characteristics.

\section{Conclusions}\label{sec:conclusions}

In this paper, a model to characterize the intermittent behaviour of battery-less LoRaWAN devices has been presented. We have analysed its performance for LoRaWAN Class A, in order to explore the limits that these devices can support. 
We have shown that the packet delivery ratio depends on device configurations (capacitor size, turn-on voltage threshold), application specification (transmission interval, packet size) and environmental conditions (energy harvesting rate). Our results showed that a realistic capacitor of $4700\mu{}F$ can support SF7 and 9 at a low energy harvesting rate of 1mW, even when transmitting more than once per minute. SF11 can only be supported for small packet sizes and high harvesting rates of 100mW or more. Finally, we showed that the device turn on threshold significantly affects performance, and it is necessary to dynamically tune it, based on environmental and application characteristics.

\section*{Acknowledgment}
Part of this research was funded by the Flemish FWO SBO S004017N IDEAL-IoT (Intelligent DEnse and Long range IoT networks) project.

\bibliographystyle{IEEEtran}
\bibliography{biblio}

\end{document}